\documentclass{iaus}
\usepackage{graphicx}

\title[Connecting Recurrent Novae to  Type Ia Supernovae] 
{Connecting Recurrent Novae\\ to [some] Type Ia Supernovae}

\author[Ferdinando Patat]   
{Ferdinando Patat$^1$
 }

\affiliation{$^1$European Southern Observatory, \\ K. Schwarzschild-Str. 2,
D-85748, Garching b. M\"unchen, Germany \\ email: {\tt fpatat@eso.org}}

\pubyear{2008}
\volume{xxx}  
\pagerange{119--126}
\jname{Title of your IAU Symposium}
\editors{A.C. Editor, B.D. Editor \& C.E. Editor, eds.}
\begin{document}

\maketitle

\begin{abstract}
In this review I summarize the observational attempts done so far to
unveil the nature of the progenitor system/s of Type Ia Supernovae. In
particular, I focus on the most recent developments that followed the
alleged detection of circumstellar material around a few events, and on
the link this possibly establishes with recurrent novae. In this
framework, I then discuss the case of RS Oph, what we know of its
circumstellar environment, and what this is telling us about its
supposed connection to Type Ia supernova explosions.


\end{abstract}

\begin{verse}
 Going after one beloved star today\\
 and after another beloved star tomorrow\\
 will not bring us very far.\\
\flushright{\cite{alvio2011}}
\end{verse}

              
\section{Introduction}

There is no need here to spell out why it is important to understand
the nature of Type I Supernovae (SNIa) progenitors. During this
conference this has been clearly and explicitly stated in many
occasions (see for instance the review by M. Livio, these proceedings).

Starting with the early seventies, when the currently accepted general
binary scenario was proposed (Whelan \& Iben \cite{wi73}), the
question mark about the nature of the donor and the accreting star has
been growing bigger and bigger. In recent years the seminal question
``single- or double-degenerate?'' has evolved, and we are now asking
ourselves to what fraction the two channels (hereafter SD and DD,
respectively) contribute to the global rate of SNIa. In the case of
the SD, we would also like to know what is the
contribution of theme's variations, where the donor star is a red
giant, a sub-giant, a MS-star, or a He-star.  The problem has become a
challenge that many groups, both theoretical and observational, have
taken.

In this review I will focus on the observational side, going through
the various attempts that have been made so far and the recent
developments. Here I am on purpose neglecting SN rate approach (see
for instance Greggio \cite{greggio10} and references therein), as this
is treated in a number of contributions in these proceedings.

\section{In quest of the prey}

So far a number of different ways of getting insights on the
progenitor's nature were explored. These are recapped in the
following sub-sections, along with their pros and cons.

\subsection{Direct observation of the progenitor system}

This obviously provides the most direct information on the system. So
far three different tracks have been beaten:

\begin{enumerate}
\item If the system is formed by a WD accreting material from a
  non-degenerate companion star, the latter is supposed to survive the
  explosion. As a consequence of the WD disruption, it gets a
  gravitational kick, and it runs away at an unusually high speed
  (when compared to all other field stars). On the contrary, if the SN
  is the outcome of the merger of 2 WDs, then there should be no
  high-speed star left over.  Since this requires multi-epoch accurate
  astrometry of the field surrounding the SN site, the method can only
  be applied to historical SNe in the Galaxy.  In particular,
  Ruiz-Lapuente et al. (\cite{pilar04}) have identified what might be
  the donor star to Tycho SN exploded in 1572. This would clearly
  testify in favor of a single-degenerate system for this specific
  event. However, the finding has been questioned by Kerzendorf et
  al. (\cite{kerzendorf09, kerzendorf11}. See also these proceedings),
  who did not identify an unambiguous donor star candidate.

\item SD progenitor systems are supposed to spend some million years
  in a phase of steady nuclear burning while accreting matter from the
  donor. In virtue of this fact, they should be detectable as
  super-soft X-ray sources (SSS) (Di Stefano
  \cite{distefano10a}). Under certain circumstances also DD systems
  may undergo a SSS phase, but with X-ray luminosities one order of
  magnitude lower than in the previous case (Yoon, Podsiadlowski \&
  Rosswog \cite{yoon07}, Di Stefano \cite{distefano10b}). Therefore,
  the detection of SSS coincident with a SN site would allow a direct
  study of [some of] the progenitor properties. This methods has been
  first applied to SN~2007on, for which pre-explosion X-ray archival
  data were available (Voss \& Nelemans \cite{voss08}). The claim of a
  detection (and the consequent conclusions on the progenitor's
  nature) was questioned by Roelofs et al. (\cite{roelofs08}).  A
  similar study was conducted on the recent, nearby
  SN~2011fe. However, only upper limits on the X-ray luminosity could
  be set, and they are not sufficiently deep to disentangle between the
  two scenarios (Li et al. \cite{li11}). The main limit of this method
  is that it requires pre-explosion archival X-ray images of the SN
  site. This problem might be mitigated by a deep X-ray survey of a
  sample of nearby galaxies.

\item In absolute terms, prior to the explosion the progenitor WD is a
  faint star. Therefore, its direct optical detection is out of
  discussion, at least for extra-galactic events. However, the
  companion star might be detectable, depending on its luminosity and
  this opens a possibility similar to the analog for core-collapse
  events (Smartt \cite{smartt09}). This was attempted for a number of
  SNIa (see Maoz \& Mannucci \cite{maoz08}, Smartt \cite{smartt11}),
  but the limits were not sufficiently stringent to rule out any of
  the possibilities. Things changed substantially with the very recent
  discovery of the nearby SN~2011fe in M101 ($d\approx$6 Mpc). Based
  on pre-explosion HST images Li et al. (\cite{li11}) were able to
  place a limit that rules out a red-giant star, and hence a symbiotic
  system like RS Oph, for this particular event. The He-star channel
  is only partially excluded, while a system with a MS-star
  transferring mass via Roche lobe overflow is fully consistent with
  the data, as is a DD configuration.
\end{enumerate}

\vspace{5mm}

The methods listed so far provide direct properties of the
progenitor. However, at least with the current observing capabilities,
they can only be applied to a small number of objects, which makes
them not really suitable for a statistical study. In the following I
discuss what can be indicated as indirect methods. They have the
advantage over the direct methods of being applicable to a larger
number of objects. However, they suffer from the fact that the
information they provide is not directly related to the progenitor,
but to what of it can be deduced from its surroundings.

\subsection{Search for entrained material}

The impact of the SN ejecta on the companion star is expected
to strip its envelope (in the case this is not a WD). Part of it
becomes entrained in the ejected material and should be observable at
late phases in the form of narrow emission lines (Wheeler et
al. \cite{wheeler75}, Fryxell \& Arnett \cite{fryxell81}, Taam \&
Fryxell \cite{taam84}, Chugai \cite{chugai86} Livne et
al. \cite{livne92}, Marietta et al. \cite{marietta00}, Meng et
al. \cite{meng07}, Pakmor et al. \cite{pakmor08}). So far, no trace of
hydrogen (or helium) has been detected in the late spectra of Type Ia
SNe and this rules out systems with secondary stars close enough to
the exploding WD to be experiencing Roche-lobe overflow at the time of
explosion (Leonard \cite{leonard07}).

\subsection{High-velocity Ca~II features in early SNIa spectra}

A feature which is common to most of SNe Ia is the presence of
high-velocity components in the Ca~II NIR triplet (Wang et
al. \cite{wang03}, Mazzali et al. \cite{mazzali05}). Gerardy et al
(\cite{gerardy04}) have explained these features as formed within a
high velocity shell produced by the interaction between the SN ejecta
and small amounts of circumstellar material (M$\leq$10$^{-2}$
M$_\odot$). The models show that this material,
which must be very close to the exploding star (i.e. within a few
10$^{15}$ cm), would not be revealed in any other form (variations of
the light curve or appearance of narrow emission lines; see also next
sub-section). The CSM material might come from the accretion disk, the
Roche lobe, or the common envelope (Gerardy et al. \cite{gerardy04}).
It must be noticed that these features can be explained by a 3D
structure of the explosion (Mazzali et al. \cite{mazzali05}) and,
therefore, circumstellar interaction is not necessarily a unique
interpretation (Quimby et al. \cite{quimby06}).

\subsection{Search for signs of ejecta-CMS interaction}

The previous method is supposed to probe the immediate surrounding of
the SN. However, in the case the donor star is a red-giant, one
expects to have a significant amount of matter at larger distances
from the progenitor. If this is the case, then the SN ejecta will
sooner or later crash into it, and some fraction of the kinetic energy
converted into optical, UV and radio emission. This was probably the
most systematically pursued channel for addressing our question.
However, all attempts of detecting direct signatures of circumstellar
material in normal SNIa using this method were frustrated, and only
upper limits to the mass loss rate could be placed from optical
(Lundqvist \cite{lundqvist03, lundqvist05}; Mattila et
al. \cite{mattila05}), radio (Panagia et al. \cite{panagia06}, Chomiuk
et al. \cite{chomiuk11}), and UV/X-Ray emission (Immler et
al. \cite{immler07}).

Assuming a typical $\rho \propto r^{-2}$, spherically symmetric,
steady red-giant wind with a velocity of 10 km s$^{-1}$, the most
stringent VLA non-detections imply mass loss rates below a few
10$^{-8}$ M$_\odot$ yr$^{-1}$ (cf. the case of SN 2006X, Stockdale et
al. \cite{stockdale06}).  Before the start of E-VLA operations, the
lack of radio detection could only exclude the symbiotic channels at
the higher hand of mass loss rates (10$^{-5}$ - 10$^{-6}$ M$_\odot$
yr$^{-1}$). However, the augmented sensitivity of E-VLA has now pushed
down this limit quite significantly. The conclusion is that either all
SN Ia observed with the refurbished VLA facility come from DD systems,
or the CSM scenario adopted so far to interpret the radio data needs
to be radically reconsidered (Chomiuk et al. \cite{chomiuk11}; see
also L. Chomiuk's contribution, these proceedings).

\subsection{Study of the SN Remnant}

The properties of young SN remnants are linked to the structure of the
medium in which they expand. In turn this is related to the material
outflows undergone by the progenitor system before it exploded (both
from the RG and from the accretion winds that are supposed to come
from the WD). Therefore, some independent information on the
progenitor can be gained looking at young SN remnants (Badenes et
al. \cite{badenes07}; Reynolds et al. \cite{reynolds07}; Chiotellis et
al. \cite{chiotelllis11}; see also the contribution by K. Long in
these proceedings). Although the application of the method is still in
its youth, and is significantly limited by the [so far] small number
of remnants that can be studied, it will certainly provide additional
constraints.

\section{CSM detection via absorption lines}

All the indirect methods presented above require a direct interaction
between the SN ejecta and the circumstellar material. But what about
cavities around the SN? If nova-like evacuation mechanisms proposed
Wood-Vasey \& Sokoloski (\cite{woodvasey06}) are at work, then the
immediate surroundings of the explosion site have all but a
spherically symmetric, $\rho \propto r^{-2}$ configuration. The
presence of an accretion disk, the bi-polar structure of the recurrent
nova (RN) ejecta and its possible fragmentation further complicate the
picture. Because of this, and especially in the early epochs of the SN
evolution, it is possible that we do not see signs of interaction
(down to the E-VLA limits) just because there is no (or not enough)
circumstellar material to interact with. In addition, if the material
is confined within a series of nested (and fragmented) shells, then
the interaction may last only a limited amount of time, hence making
its detection via radio and/or X-ray emission more difficult.

With these speculative considerations in mind, back in 2006 my
collaborators and I started considering an alternative approach to the
problem, which would allow the detection of small amounts of material
without requiring direct interaction with the SN ejecta. This
technique and its first application is presented in Patat et
al. (\cite{patat07}), and further expanded in Simon et
al. (\cite{simon09}) to which I refer the reader for a more
quantitative treatment. The idea is based on the fact that, at
variance with core-collapse SN, Type Ia are relatively weak UV
sources, because of strong line-blocking by heavy elements
(Fe/Co/Ti/Cr; Pauldrach et al. \cite{adi}; Mazzali et
al. \cite{mazzali00}). As a consequence, their ability to ionize
possible circumstellar matter is rather limited in
distance. Therefore, provided that the gas is placed along the line of
sight, it can be revealed by optical absorption lines like Ca~II H\&K,
Na~I D, and K~I. These are all strong lines, so intense that they
enable the detection of tiny amounts of material. As opposed to the
emission lines that are expected to arise only in the case of direct
ejecta-CSM interaction (mainly H and He), absorption lines are
generated by material placed virtually at any distance along the line
of sight (and only along the line of sight, which is one of the
drawbacks of the method).

Just because of this (one may object), there is in principle no
difference between a line arising in the CSM and one generated within
a ISM cloud placed along the line of sight. However, given its weak
ionizing radiation field, a SN Ia is hardly capable of ionizing gas
placed at distances larger than $\sim$10$^{18}$ cm. And even if it
were, the typical ISM densities would imply very long recombination
times, meaning that no time variations are expected in lines arising
in the ISM. On the contrary, this may not be true for absorptions
produced by dense gas close to the explosion site. If the density is
high enough, the material would be initially (and possibly partially)
ionized by the SN radiation field. Then, as the SN fades away, the
material recombines, and the lines get more and more intense as time
goes by. Obviously, at some point this material may also be reached by
the SN ejecta. Although its mass could be insufficient to generate a
measurable emission (and would therefore make any X-ray and radio
detection fail), the interaction would re-ionize the material, causing
the absorption features to disappear. Given that the SN ejecta
velocity is known, one can deduce the distance of the material from
the SN just from the timing of the line disappearance. The SN ejecta
move at typical velocities of 3$\times$10$^{4}$ km s$^{-1}$. As a SNIa
rises to maximum light in about 20 days, this implies that at this
epoch the ejecta have traveled $\approx$6$\times$10$^{15}$ cm
($\sim$400 AU). This gives an idea of the spatial scales that can be
probed by this method.

These ideas were first applied to the bright SN~2006X, for which we
obtained a series of high resolution spectra with the VLT+UVES. When
we compared the first two epochs (separated by two weeks around
maximum light) we found the line profile of Na~I D had profoundly
changed. On the blue side of a deep absorption (which was easily
identified as the signature of a thick molecular cloud associated with
the local spiral arm) a series of weak features had appeared,
indicating the presence of material moving towards the observer at a
velocity between 50 and 100 km s$^{-1}$ (Patat et al. \cite{patat07};
Patat \cite{patat09}). The conclusion was that these features were
originating either in the wind of a red giant (RG), or in the remnant
shells of successive nova outbursts. Both possibilities favored the SD
scenario (but see footnote 1 in Patat \cite{patat09}).

A second variability detection was reported for SN~2007le by Simon et
al. (\cite{simon09}), who reached similar conclusions about the
circumstellar nature of the material generating the variable Na~I D
absorption. As in the case of SN~2006X, no time evolution was seen in
the Ca~II H\&K profile, strongly testifying in favor of a radiation
effect (as opposed to a geometrical effect; see Chugai \cite{chugai08}
and Patat et al. \cite{patat10}). As of today, thirteen SN Ia were
studied by means of multi-epoch, high-resolution spectroscopy (Cox et
al. \cite{cox11}, Patat et al. \cite{patat11b}, Simon et
al. \cite{simon11}). Out of these, only two have shown line
variability\footnote{Another two cases of line variability were
  reported for SN~1999cl (Blondin et al. \cite{blondin09}) and
  SN~2006dd (Stritzinger et al. \cite{stritzinger10}). However, they
  are both based on low-resolution data, and not much can be said
  about the physical conditions of the gas in which the lines
  originate.}. This can be interpreted in terms of viewing angle
effects, the presence of other SD sub-channels, or even DD. Although
the observations are still in progress while this review is being
written, I can anticipate that the high-resolution data obtained so
far for the nearby SN~2011fe do not show any trace of Na~I D
variation. It is worth noting that for this one object we have the
most stringent limit on the mass loss rate: prompt radio and X-Ray
observations have yielded $dM/dt\leq$10$^{-8}$ (w/100 km s$^{-1}$)
M$_\odot$ yr$^{-1}$ (Horesh et al. \cite{horesh11}). Moreover, we know
the progenitor is not in a symbiotic system that would undergo a RN
phase (Li et al. \cite{li11}), so that no complex CSM structure is
expected. Of course, SN~2011fe is just one object and it does not tell
us how frequent these systems are. For instance, one might ask the
question whether the frequency of events displaying Na~I variations is
related to the occurrence of a particular progenitor channel (more
specifically the symbiotic channel). Although we are still far from
being able to answer this question, some statistical evidence is
emerging. The recent study by Sternberg et al. (\cite{assaf11}) has
shown a significant excess of blue-shifted Na~I features in SN Ia,
which are interpreted as a systematic sign of gas outflows at
velocities ranging from a few 10 km s$^{-1}$ to more than 100 km
s$^{-1}$. These indicate that the SD channel (with possible
sub-channels) dominates over the DD channel.

One issue with the conclusions drawn from the line variability studies
is that we are not yet sure RNe do indeed generate structures in the
CSM that i) have the required velocities, and ii) survive long enough
to reach the required distances ($>$10$^{16}$ cm). At this point one
needs to close the loop, asking oneself the question: do RN systems
show similar features?

\begin{figure}[t]
\vspace*{0.0 cm}
\begin{center}
\includegraphics[width=3.4in]{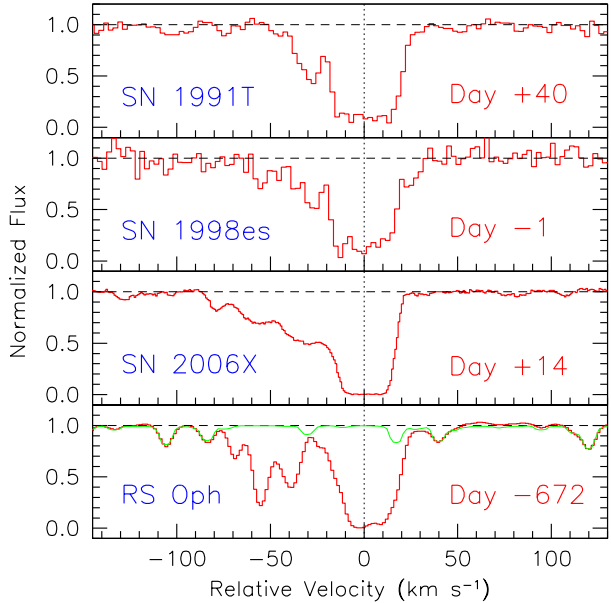} 
\vspace*{-0.0 cm}
\caption{Comparison between high-resolution spectroscopy of three SN
  Ia and RS Oph in the region of Na~I D$_2$. For presentation the
  velocity scales refer to the center of the deep absorption, most
  likely associated to the local spiral arm. The light colored curve
  in the bottom panel is an atmospheric model.}
\label{fig1}
\end{center}
\end{figure}

\section{Closing the loop: the case of RS Oph}

RS Oph is composed by a hot component that accretes material from a
RG, and that it undergoes RN outbursts with a period of roughly 20
years, the last of which took place in February 2006 (see the reviews
by Mikolajewska and Anupama in these proceedings).  The hot component
is identified as a WD close to the Chandrasekhar mass. For this
reason, RS Oph has been proposed as a viable candidate to explode as a
SNIa, and it therefore constitutes a most suitable laboratory to
address the above question. RNe offer the possibility of being studied
before, during, and after the outburst, and this is a great advantage
over SNe.  We profited from this property studying multi-epoch,
high-resolution data taken across the 2006 outburst.  The analysis is
rather complex and the details can be found in Patat et
al. (\cite{patat11}). The main conclusion is that a number of
blue-shifted Ca~II, Na~I, and K~I absorption lines, with velocities
reaching 50 km s$^{-1}$, are associated to material not belonging to
the circumbinary environment. The resemblance to what is seen in some
SN Ia is striking (Fig.~\ref{fig1}).  These features, which during the
outburst tend to weaken (or even disappear), show up again in the data
taken about two years after the outburst, demonstrating that the gas
in which they originate have survived the eruption, and must therefore
be placed at distances larger than $\approx$3$\times$10$^{15}$ cm. The
most natural explanation for these features is that they are generated
in structures produced by the interaction between the previous nova
shells and the RG wind lost in the inter-outburst phase.  The
relatively low velocities observed for these features require that the
nova shell is significantly decelerated while expanding in the
material lost by the RG. Is this physically possible within the
framework of known facts about RN systems like RS Oph?

Assuming spherical symmetry and momentum conservation, and using the
known parameters of the system (ejected mass, shell kinetic energy and
initial velocity, inter-outburst period, RG wind velocity), one can
see that in order to decelerate the nova shell a mass loss rate of
3$\times$10$^{-7}$ M$_\odot$ yr$^{-1}$ is required (Patat et
al. \cite{patat11}). Despite the crude simplifications, this value is
fairly consistent with what is expected for a RG in a symbiotic
system: during the inter-outburst phase the RG loses enough material
to slow down the next nova shell at velocities below 100 km
s$^{-1}$. Therefore, subsequent eruptions produce a series of nested
structures, which would not physically interact with each other, but
would rather remain separated in radius by $\sim$10$^{15}$ cm.
Although this conclusion lends support to the proposed scenario, the
simplified picture neglects two important facts: a) the nova ejecta
deviate from spherical symmetry (O'Brien et al. \cite{obrien06},
Sokoloski et al. \cite{sokoloski06}, Ribeiro et al. \cite{ribeiro09}),
and b) Rayleigh-Taylor instabilities lead to shell fragmentation. The
hydrodynamical simulations presented during this conference by
S. Mohamed (see these proceedings) have shown what kind of structures
one is to expect and what the viewing angle effects would be. Indeed,
the required structures, densities and filling factors seem to be
consistent with the above picture\footnote{Dr Mohamed, I love you.}.
 The remaining question is whether these structures would
survive long enough to reach 10$^{16}$-10$^{17}$ cm, to make them
viable for the SN case (SN ejecta are a factor ten faster than in
novae).

A number of questions remain unanswered. For instance, it is not yet
clear whether the hot component in RS Oph is indeed a C-O or rather a
He-Ne-Ng WD. This would of course make a big difference. Another
relevant point to be addressed is whether what we have seen in RS Oph
is common to all known RN systems. This will require a more systematic
study of these systems, and may tell us whether the presence of
narrow, blue-shifted absorption lines can be used to identify SN Ia
coming from this channel. Another fundamental piece of information we
need is the survival time of the knots produced by the nova shell
fragmentation. Future hydrodynamical simulations along the lines
presented by Mohamed will have to extend over many nova outbursts to
see whether these knots can reach the distances required to show up in
SNe. Also, they will give us more quantitative estimates about the
viewing angle effects, and what is to be expected when observing SN
Ia.

\vspace{3mm}

Alvio Renzini is right, and the matter about SN Ia progenitors needs
to be settled within the wider context of binary stellar
evolution. But I believe that there is something to learn from single
objects, like RS Oph, too (maybe just to reach the conclusion that
such objects cannot explode as SN Ia). It is probably the synergy
between the two that will lead us somewhere. All the pieces that at
times (at least to me) seem just a contradictory set of information
(as W. Hillebrandt once said half serious half joking, "everything
taken into account, Type Ia SN should never take place") may then fall
together to form a comprehensive picture.

\end{document}